\documentclass[]{interact}

\usepackage{epstopdf}
\usepackage[caption=false]{subfig}

\usepackage[numbers,sort&compress]{natbib}
\bibpunct[, ]{[}{]}{,}{n}{,}{,}
\renewcommand\bibfont{\fontsize{10}{12}\selectfont}

\usepackage{ulem}
\usepackage{xcolor}

\definecolor{MyGreen}{RGB}{0, 128, 0}

\usepackage{amsmath}

\makeatletter
\def\NAT@def@citea{\def\@citea{\NAT@separator}}
\makeatother

\theoremstyle{plain}

\theoremstyle{definition}

\theoremstyle{remark}

\begin{document}
	
	
	
	\title{CONSENSUS IN ASYNCHRONY: STRICTLY FORMAL}
	
	\author{
		\name{Ivan Klianev \thanks{CONTACT Ivan Klianev. Email: Ivan.Klianev@gmail.com}}
		\affil{Transactum Pty Ltd, Sydney, NSW, Australia}
	}
	
	\maketitle
	
	\begin{abstract}
		In this paper, we resolve the apparent contradiction between the proven possibility of deterministic crash-tolerant consensus in a fully asynchronous environment and the reconfirmation of the FLP impossibility result by Attiya, Castañeda, and Rajsbaum. With the use of a strictly formal framework that extends their reasoning, we close this fundamental gap in theory. Specifically, we demonstrate that a single protocol phase separates their findings from reaching the exact opposite conclusion. Another important outcome is a novel algorithm with ability to tolerate multiple crash faults.	We provide a rigorous, strictly formal proof of correctness to validate our results.
	\end{abstract}
		
	
	\section{INTRODUCTION}
	
		Two distinct types of gaps are associated with theoretical sciences. The first concerns the divergence between mathematical models and their real‑world implementations. The other arises when a mathematically correct model offers a narrow or even distorted representation of reality. The gap between theory and practice reflects the idea that we cannot yet execute perfectly, whereas the gap between theory and reality reflects the deeper issue that we do not yet understand perfectly. Computer science, like any other theoretical discipline, is also not immune to treating preliminary or incomplete results as exhaustive and final. This work closes a fundamental gap in computer science.
		
		We resolve the apparent contradiction between our demonstrated possibility \cite{Klianev_2026} of deterministic crash-tolerant consensus in completely asynchronous systems and the reconfirmation of the FLP impossibility result \cite{FLP_result} by Attiya, Castañeda, and Rajsbaum (Attiya et al.) \cite{AttiyaEtAl_2023}. Specifically, we demonstrate that a single missing protocol phase separates their findings from reaching the exact opposite conclusion.
		
		We adopt the same \textit{shared‑memory} computational model used by Attiya et al. Our analytical framework draws on a mathematical structure from order and lattice theory \cite{DaveyAndPriestley}, combining the key notions of \textit{join‑semilattices} and \textit{monotonicity}. By examining the event‑based synchronization implemented by the consensus algorithm, we demonstrate not only the tolerance to multiple crash faults but also the tolerance to more crashes relative to the number of processes than the previously shown in our earlier work.
		
	\subsection{\textsf{Background}}
	
	\subsubsection{\textsf{Local Proof of Impossibility}}
	
		The FLP theorem identifies a natural constraint: in a fully asynchronous system, termination of a crash‑tolerant consensus protocol is impossible when processes rely solely on exchanging \textit{initial input} messages. This constraint holds regardless of whether the agreement task is data‑dependent or data‑independent.
	
		The same constraint is confirmed by the CALM theorem \cite{Hellerstein_and_Alvaro_2020}. It shows that in the absence of faults, no consensus protocol is required. The exchange of initial inputs does not itself constitute a consensus protocol.  It is the preliminary distribution that precedes the unavoidable need for a true consensus mechanism once faults are possible.
		
		When the absence of such a mechanism threatens consensus safety, the FLP proof employs the valency argument to show impossibility to terminate without sacrificing safety. In effect, it interprets the emergence of need for a consensus protocol as evidence of inherent impossibility of crash‑tolerant consensus in an asynchronous setting.
		
		The valency argument is fundamentally \textit{local}: the FLP theorem reasons from a small fragment of the global state, specifically, the initial inputs known to a single process. Yet this local reasoning implicitly establishes a broader theoretical insight. It shows the impossibility of \textit{any} agreement in full asynchrony using a crash‑tolerant consensus algorithm that operates in a single phase and exchanges only initial input messages.
		
	\subsubsection{\textsf{Global Reconfirmation of Impossibility}}
	
		Our earlier work reconciled its results with the \textit{local} proof, but not with the global‑state argument by Attiya et al. Their analysis clarifies the valency argument limits and reconfirms that crash‑tolerant consensus in asynchrony has no locally solvable path: it fundamentally requires global consistency and therefore demands a global proof.	
	
		Their work examines what processes can achieve by exchanging views of the global state. It shows that consensus in asynchronous systems with failures is neither locally solvable nor solvable through the exchange of global‑state views. It strengthens the impossibility result: even after processes share their views of the global state, any attempt to enforce a global decision still violates safety.

		Practically, exchanging global‑state views corresponds to exchanging vectors of known initial inputs, which may later be used to compute a binary agreement value. The focus is not on the processes’ inability to compute a binary value due to the data configuration. It shifts to how computing a binary agreement value from the initial inputs of different sets of processes can itself lead to a safety violation. 
		
		Their work establishes impossibility that is stronger and more convincing than the one derived from the valency argument. It shows that consensus remains impossible even with guaranteed termination through vector agreement, after processes exchange both their initial inputs and their global‑state views. More broadly, it demonstrates impossibility with \textit{any} consensus algorithm that operates in \textit{two phases}.

	\subsubsection{\textsf{Pending Reconciliation}}

		We reconcile the possibility of solving consensus in \textit{three phases} with the impossibility of doing so in \textit{two phases}. Our framework combines the \textit{shared‑memory} model with a monotonic \textit{\textit{join‑semilattice structure}}. Using it, we demonstrate the impossibility of consensus in two phases, the possibility in three phases, and the precise way in which the \textit{third} phase builds upon the foundational work completed in the \textit{second}.

	\subsection{\textsf{Motivation}}
	
		Fault‑tolerant consensus in complete asynchrony enables non‑blocking distributed transactions \cite{Gray_Lamport} and active‑active replication with both consistency and availability. Compared with consensus in partial synchrony, it operates under fewer assumptions.
		
		A typical assumption of the fully asynchronous model – the absence of faulty links – can be relaxed. As a result, the system must instead operate within a set of bounded ratios relating tolerated faulty links to actual crash failures, which can be computed straightforwardly. Under this single remaining condition, consensus ensures both safety and liveness, and systems built on top of it provide both consistency and availability.
	
	\subsubsection{\textsf{Active-Active Replication}}

		Active‑active replication is a setting in which binary agreement is irrelevant;
		to impose a sequential order on incoming transaction requests it requires vector agreement. To preserve both consistency and availability in a system with three or more replicas, every replica in a minority partition, isolated from the majority due to asynchrony, must still be able to serve requests without jeopardising the consistency of the replicated state.
	
		The algorithm for consensus in asynchrony addresses this challenge, formalised by the CAP theorem \cite{GilbertLynch_2002}, in the following way. It guarantees that any process whose initial input is not included in the agreed vector nevertheless terminates with the same agreed vector as all other correct processes. From a replication perspective, this ensures that partitioned replicas execute all transaction requests contained in the agreed vector, while refraining from executing the requests presented only in their own initial inputs. Those omitted inputs will be incorporated by the majority in a subsequent consensus round, at which point the corresponding transactions will be executed by all replicas.

	\subsubsection{\textsf{Distributed Transactions}}

		Distributed transactions are a setting in which binary agreement is unavoidable, and they expose four subtle but essential aspects of binary agreement in this context. Consider a system A (three replicas, active–active) executing a distributed transaction with a similar system B. In this setting:

	\begin{itemize}
		
		\item Faulty or slow replicas \textbf{do not block} progress. A consensus algorithm that tolerates two faults in a six‑process system always terminates, ensuring that the transaction completes even when some replicas are slow or faulty.		
		
		\item \textit{Ties} \textbf{are multi-faceted}. To support fine‑grained decision‑making, the binary outcome should be derived from a vector agreement, not from a naïve majority rule. Vector agreements preserve the association between each vote and its voter.
			
		\item \textit{Ties} \textbf{do not prevent} agreement. A tie does not imply indecision. If all replicas of B vote 0 (Abort) and all replicas of A vote 1 (Commit), the transaction outcome is 0 (Abort). Further, if A produces two votes for 1 and one for 0, while B produces one vote for 1 and two for 0, the outcome is 1 (Commit) under Commit-biased rule, or 0 (Abort) under Abort-biased rule.
		

		\item A \textit{tie} \textbf{is not a simple} symmetry. Its meaning depends on which replica cast which vote – information preserved only in the agreed vector.
		
	\end{itemize}
	
		These features allow to resolve the long‑standing challenge of designing non‑blocking distributed transactions \cite{BernsteinHadzilacosGoodman}. They enable the design of a fault‑tolerant integration layer capable of linking banks and financial institutions into a unified global system supporting instant fund transfers and instant delivery‑versus‑payment settlement.

	\subsection{\textsf{Problem Formulation}}
	
		Create a formal framework and use it to demonstrate in a strictly formal manner that the processes agree on an $n$-dimensional vector despite timing uncertainty and failures, by using a \textit{three-phase} algorithm, but cannot agree using a \textit{two-phase} algorithm.
		
	\subsubsection{\textsf{System State Space}}

		Let $\Pi = \{p_1, p_2, ..., p_n\}$ be a set of $n$ processes. Each process $p_i$ is provided with an initial input $v_i \in \mathcal{V}$, where $\mathcal{V} = \{0,1\}$. The goal is to decide on an output vector $\Vec{V}= \{v_1, v_2, ..., v_n \}$, where each $v_i \in \mathcal{V} \cup {\perp} $. The symbol ${\perp}$ represents a \textit{null value}, indicating the input of process, which cannot be retrieved. 

	\subsubsection{\textsf{Formal Constraints}}

		For a vector consensus algorithm to be correct in total asynchrony with $f$ crashes, it must satisfy three mathematical properties:

		\textbf{- Agreement (safety)}

		If two correct processes $p_i$ and $p_j$ decide on vectors $\Vec{Y_i}$ and $\Vec{Y_j}$ respectively, then:
		\begin{equation*}
			\Vec{V_i} = \Vec{V_j}
		\end{equation*}

		\textit{In short}: Every process that has decided must see the exact same vector.

		\textbf{- Validity (integrity)}

		The decided vector $\Vec{V}$ must satisfy:

		a) Input preservation: For at least $(n-f)$ indexes $k$, $v_k \in \{v_k, {\perp} \}$.

		b) Fault limit: The number of \textit{null values} is bounded by the number of crashes:
		\begin{equation*}
			| \{ k : v_k = {\perp} \} | \le f
		\end{equation*}

		\textit{In short}: The vector must be composed of real inputs sent by the processes, with at most $f$ missing.

		\textbf{- Termination (liveness)}

		For every correct process $p_i$, there exists a finite time $t$ such that $p_i$ enters decision state:
		\begin{equation*}
			\forall p_i \in \Pi_{correct}, \exists t < \infty : decide (p_i, \Vec{V_i} )
		\end{equation*}

	\subsubsection{\textsf{Problem Solving Condition}}

		The problem is solved when it is demonstrated that:
		
		1 A set of processes $S$, $S \subset \Pi$, which is the set of processes whose input values are received and included in the output vector $\Vec{V}$ satisfies the \textit{quorum condition}:
		\begin{equation*}
			| S | \ge (n-f)
		\end{equation*}
		
		2. Only a consensus algorithm that executes in \textbf{no less} than \textit{three phases} guarantees that $\forall p_i \in \Pi_{correct}$ and $\forall p_j \in \Pi_{correct}$:
		\begin{equation*}
			\Vec{V_i} = \Vec{V_j}
		\end{equation*}

	\subsection{\textsf{Research Question}}	
	
	Atomic broadcast \cite{Rodrigues_and_Raynal_2000} is a foundational challenge in distributed systems: it ensures that all correct processes deliver the same set of messages in exactly the same order, even in the presence of failures. It combines reliable delivery with strict total ordering. 

	Processes to agree on an $n$-dimensional vector, despite the timing uncertainty and failures, requires ability to transform reliable broadcast into atomic broadcast: i) with no leader, physical clock, or synchrony assumptions; and ii) in spite of crashes.
	
	\textit{The question}: If a deterministic algorithm that produces a consistent linear extension of a distributed partial order \cite{Chan_and_Pak_2025} exist, can it operate in less than \textit{three phase}s?
	
	\subsubsection{Formulation: Input and Output}
	
	We describe the conceptual challenge as a mapping between two types of ordered sets:
	
	\vspace{\baselineskip}
	
	\textbf{Input: Reliable Broadcast (RB)}
	
	A set of messages $M$ delivered by processes in \textit{partial order} $(M, \prec_{RB})$:
	
	\begin{itemize}
		\item In an asynchronous system with crashes, different processes, $p_i$ and $p_j$, may observe messages in different local arrival sequences.
		
		\item The only inherent order possible to implement is \textit{causal order}, where $m_1 \prec m_2$ if $m_1$ happened before $m_2$. Otherwise, messages are concurrent and incomparable.
	\end{itemize}
	
	\textbf{Output: Atomic Broadcast (AB)}
	
		A \textit{totally ordered} set $(M', <_{AB})$ such that for any two messages, $m$ and $m'$, every correct process $p_i$ agrees on whether $m <_{AB} m'$ or $m' <_{AB} m$.
		
	\subsubsection{\textsf{Design Problem: Algorithm with Higher Tolerance To Crashes}}
	
		The previous work \cite{Klianev_2026} presents a function for deterministic linearisation $\Phi$, which maximises the inclusion of the initial inputs in the agreed vector in spite of asynchrony and a crash. It demonstrates ability of a 5-process system to tolerate only 1 crash. 
	
		Now, the design objective is function $\Phi$ to tolerate multiple crashes and more crashes per number of system processes. Thus, a system of 5 processes must tolerate 2 crashes.
	
	\subsubsection{\textsf{Research Problem: Deterministic Linearization}}	
	
		Strictly formal, the research problem is to conceptually define a set of requirements that a deterministic function $\Phi$ must satisfy in order to transform the observed partial order into a total order without external linearization:
		\begin{equation*}
			\Phi : (M, \prec_{RB}) \rightarrow (M', <_{AB})
		\end{equation*}
	
	\subsubsection{\textsf{Analysis and Explanation}}
	
		On condition of demonstrated existence of algorithm with the required performance characteristics, use the formal framework to explain:
		
		1. Why it is impossible to terminate with guaranteed safety in \textit{two phases}.
		
		2. How it becomes possible to terminate with guaranteed safety in \textit{three phases}.
	
	\subsection{\textsf{Contributions}}	
	
		The primary contribution of this paper is the resolving of a critical gap in the theoretical foundation of asynchronous consensus. Specifically, our core contributions are threefold:
		
	\begin{itemize}
		\item \textbf{Identification of the Analytical Gap}: We demonstrate that the impossibility analysis by Attiya et al. stopped exactly one protocol phase short of an opposite conclusion, thereby clarifying the precise boundary between impossibility and possibility in deterministic consensus.
		
		\item \textbf{Novel Asynchronous Protocol}: We introduce a novel deterministic consensus protocol designed for fully asynchronous networks that successfully tolerates multiple crash faults.
		
		\item \textbf{Rigorous Formal Verification}: We provide a strictly formal proof of safety, validity, and liveness for our protocol, mathematically establishing its validity and resolving the apparent contradiction with established impossibility results
		
	\end{itemize}
	\section{\textsf{RESEARCH METHODS}}	

	\subsection{\textsf{System Model}}

		A locks-free shared-memory system $\Pi $ with $ n \ge 3$ asynchronous processes, $\Pi = \{p_1, p_2, ..., p_n\}$ uses inter-process communication model, where the processes access a common memory segment for reading and writing data. The id of process $P_i$ is $i$. Processes may fail by crashing, i.e., stop taking steps. The links never fail. The system is specified by a deterministic transition function $\Phi$ associated with each of the processes and their individual state. The processes strictly follow the transition function, i.e., there is no Byzantine behaviour.
		
	\subsubsection{\textsf{Modelling Asynchrony}}

		Communication via shared memory adequately models two aspects of reliable broadcast over healthy asynchronous links: i) messages are never lost; and ii) the ensured \textit{all or nothing} (atomic) delivery of messages. It does not reflect two critical possibilities in asynchrony. First, the processes may receive a pair of messages from the same sender in a sequential order that is different from the order of sending. Second, it does not reflect the random order of receiving the messages broadcast by different processes. Yet, these possibilities are taken in consideration and handled by the algorithm.
		
		Writing in the shared memory is not synchronised. A process writes messages in its own dedicated area; it cannot write in other processes' areas. It cannot delete or overwrite the messages written by other processes or its own messages. Once a message is written it stays in the shared memory until all non-crashed processes terminate.
		
		A process writes in the shared memory when the transition function $\Phi$ obligates it to broadcast a message. Apart from the initial broadcast, every next broadcast is in response to an event of received message. The processes execute in parallel on individual CPU cores. For this reason, reading from an area in the shared memory is synchronised with writing in the same area in a sense that reading of a particular area cannot be executed before completion of writing in that area.

	\subsubsection{\textsf{Modelling Crashes}}

		A process may crash at any point during execution of a consensus algorithm. A crashed process stops executing steps of the state transition function $\Phi$. This means that it no longer reads messages. A process may crash in between two writing in the shared memory; never during the writing.
		
	\subsubsection{\textsf{State Transitions}}

		Each process $p_i$ starts with an initial input $v_i \in \mathcal{V}$, where $\mathcal{V} = \{0,1\}$. The goal is to decide on an output vector $\Vec{V}= \{v_1, v_2, ..., v_n \}$, where each $v_i \in \mathcal{V} \cup {\perp} $. The number of elements in $\Vec{V}$ having $\perp$ value must be $\le f$, where $f$ is the number of tolerated crashes of processes. The state transition function $\Phi$ transforms a process's current state into a new state based on observed events: $\Phi : V_i \times E_j \rightarrow V'_i$, where:

	\begin{itemize}
	
		\item $E_j$ is the event of receiving a message $m$ containing another process's view $v_j$.
		\item $V_i$ is the state of $p_i$ before event $E_j$.
		\item $V'_i$ is the updated state of $p_i$ as a result of event $E_j$.
		\item $\Phi(v_l,m)$ is the update of state, computed as $v_i \cup v_j$.
	
	\end{itemize}

	\subsubsection{\textsf{Messages}}

		A \textit{message} is a tuple (\textit{i}, \textit{name}, \textit{number}, $m$), where $i$ is the index of the message originator process,  the \textit{name} is a 'message name' from a fixed universe of names, \textit{number} is a 'sequential number', and $ m $ is a 'message value' from a fixed universe of message values.

	\subsection{\textsf{Applicable Concepts}}
	
	\subsubsection{\textsf{Monotonic Join-Semilattice}}
	
		A mathematical structure from order theory and lattice theory \cite{DaveyAndPriestley} that combines two key ideas:
	
	\begin{itemize}
		\item \textbf{Join-semilattice}:
		\begin{itemize}
			\item A partially ordered set $P$ in which any two elements $a, b \in P$ have a least upper bound, denoted as $a \lor b$.
			\item Formally: i) Idempotent: $a \lor a = a$ ; ii) Commutative: $a \lor b = b \lor a$ ; and iii) Associative: $(a \lor b) \lor c = a \lor (b \lor c)$.	
			\item The join operation $\lor$ is order preserving in each argument.
		\end{itemize}
	\vspace{\baselineskip}
		\item \textbf{Monotonicity}:
		\begin{itemize}
			\item A function $\phi: P \rightarrow Q $ between two partially ordered sets is monotonic, i.e., order preserving, if $a \le b \Rightarrow \phi(a) \le \phi(b)$.
			\item In the context of join semilattice, a monotonic join homomorphism as a function that is both monotonic \textit{and} preserves joins: $\phi( a \lor b ) = \phi(a) \lor \phi(b)$.
		\end{itemize}
	\end{itemize}
	
	\subsubsection{\textsf{Coordination-Free Consistency}}
	
		Monotonicity, being a property of operations that ensures a non-decreasing progression over time, can facilitate consistent, coordination-free system behaviour, which means that distributed consensus is not needed. The CALM (consistency as logical monotonicity) theorem \cite{Hellerstein_and_Alvaro_2020} states that a problem is coordination-free if and only if it is monotonic, where monotonicity is defined on sets of inputs and outputs: $P(S) \subseteq (T)$ whenever $S \subseteq T$, where $P$ is the logical problem specification.	
		
	\subsection{\textsf{Algebraic Modelling of the System Model of Computation}}
	
		Compatibility between the lock‑free shared‑memory model and the monotonic join‑semilattice structure is a foundational principle in the design of coordination‑free, fault‑tolerant distributed systems. In this setting, compatibility refers to how the algebraic properties of the lattice enable shared state to be updated and merged across processes without relying on traditional synchronization mechanisms such as locks.

		A monotonic join‑semilattice can be understood as an algebraic abstraction of a lock‑free shared‑memory computational model. Conversely, a lock‑free shared‑memory model can be viewed as a concrete operational realization of the algebra defined by a monotonic join‑semilattice. The strong correspondence between these two perspectives has led to a well‑established harmony: each models the other in a way that preserves the consistency of the outcomes they independently produce.
		
		For a transition function $\delta$ to be valid within a monotonic join-semilattice, it must satisfy three formal conditions:
	
	\begin{itemize}
		\item Inflationary (monotonicity). The process never 'unlearns' information. Its state only moves upward or stays the same in the lattice, i.e., for any view $v$ and any message $m$, function $\delta$ must operate in the following manner $v \le \delta(v,m)$.
		
		\item Order preserving. If one process has a more advanced view than another, receiving the new information maintains that relative lead, i.e., if $v_1 \le v_2$, then $\delta(v_1, m) \le \delta (v_2,m)$.
		
		\item Convergence (join-equivalence). Because the underlying operation is the lattice join $(\cup)$, the final state is independent of the order in which messages $m_1$ and $m_2$ are processed: 
		$\delta(\delta(v, m_1), m_2) = \delta(\delta(v, m_2), m_1)$.
	\end{itemize}
	
		The transition function $\Phi$ of the system model fully satisfies the required conditions for $\delta$. Consequently, modelling the problem within the framework of a monotonic join‑semilattice is entirely appropriate for our purposes. 
		
		We present the problem, its solution, and the proofs using the concepts and the terminology of monotonic join‑semilattices. This allows us a rigorous formalisation, improved conceptual clarity, and proofs that are more structured and easier to follow.
	
	\subsection{\textsf{Research Through Design}}
	
		We investigated whether consensus with vector agreement can be achieved by designing \textit{two‑phase} algorithms that collect and analyse global state. None of these designs succeeded: each one of the designed algorithm eventually encountered a configuration in which the system either waited indefinitely or violated safety.
	
		Across the range of designed algorithms we verified, a recurring pattern emerged: whenever safety was violated by terminating with more than one decision vector, the set of initial inputs in the vector of smaller cardinality was a subset of the set of initial inputs in the vector of larger cardinality. Once we became confident that all possible designs of \textit{two‑phase} algorithm have been explored and their impossibility to solve consensus confirmed, this observation suggested a \textit{three‑phase} algorithm that does

		\section{\textsf{THE CONCEPT}}
	
		\subsection{\textsf{Vector Agreement}}	
	
			\textbf{Definition 1}: \textit{Vector Agreement.} Agreement where $ n $ processes $p_i$, $i\in\{1,...,n\}$: a) start with individual initial value $v_i$ that can be a vector with finite number of elements; b) individually compose an $ n $-vector $V_i=(v_1,...,v_n)$; c) terminate when $\ge (n-f)$ processes $p_i$ agree on a vector $V=(v_1,...,v_n)$ containing at most $f$ empty vector elements; and d) satisfies the properties:
	
	
		\begin{itemize}
			\item Agreement: All correct processes agree on the same vector;
			\item Validity: The agreed-upon vector contains $\ge (n-f)$ initial values as vector elements;
			\item Termination: Every correct process reaches a decision.
		\end{itemize}
	
			In pursuit of mapping between two types of ordered sets – a set of messages delivered by processes in partial order $M$ into a totally ordered set $M'$ – the \textit{union of two vectors} is necessarily treated as a \textit{disjoint union} (more accurately, a \textit{union of disjointly indexed elements}) because the information in the vector is positionally anchored to its source. This is necessary for three  reasons: i) proof of origin; ii) prevented collision of data; and iii) functional significance.
	
	
		\subsubsection{\textsf{Index Mapping as Proof of Origin}}

			In a standard  set union $(A \cup B)$, elements lose their identity regarding who provided them. In our solution's vector agreement, the vector V is a mapping from a set $\mathbb{I}$ of process IDs to a set of values $\mathcal{V}$, i.e., $V \subseteq \mathbb{I} \times \mathcal{V}$. Because each index $k$ in the vector corresponds to a unique process $p_k$, the information at $V[k]$ can only be provided by $p_k$. Thus, the sets of information being joined from different indexes are \textit{disjoint} by definition. 
	
	
		\subsubsection{\textsf{Preventing Data Collision}}

			In an asynchronous model, if two processes $p_i$ and $p_j$ were to fill the same slot, a standard union would result in a set of two values $\{x_i,x_j\}$, breaking the property of \textit{vector agreement} which requires a single value per slot. By modelling the \textit{join} as a \textit{disjoint union of mappings}:
	
		\begin{itemize}   
			\item The values are stored as pairs: \textit{(id, value)}.
		
			\item  Since $id_i \ne id_j$, the input from $p_i$ and $p_j$ occupy disjoint logical space in the resulting state.
		
			\item  Join operation $(\sqcup)$ remains idempotent – there is no ambiguity about where a value belongs.
		\end{itemize}
	
	
		\subsubsection{\textsf{Preventing Data Collision}}

			Treating the union as disjoint allows the rules for completion of algorithm phases to function. If all received values are simply merged into a single set, the algorithm could not verify whether it has received messages from $(n-f)$ distinct processes by counting the cardinality of non-${\perp}$ elements of a vector.
	
		\subsection{\textsf{Vector Agreement as Total Ordered Set}}	
		
			We reformulate the consensus problem by shifting the focus from \textbf{temporal state} to the \textbf{order theory} of messages. We treat the execution of an asynchronous system as a problem of constructing a consistent across processes order of events. Here is the formal definition of the system model using the terminology of ordered sets: 
		
		
		\subsubsection{\textsf{Message Universe as an Ordered Set}}
		
			Let $M$ be the set of all possible messages. The execution is viewed as a \textit{partially ordered} set (\textit{poset}) $(M, \le)$, where the relation $\le$ represent the causal dependency (Lamport's \textit{happen-before} relation):
		
		\begin{itemize}
			\item $m_i \le m_j$ if the receipt of $m_i$ is a necessary condition for the emission of $m_j$.
			
			\item $m_i$ and $m_j$ are not related $(m_i \parallel m_j)$ if they are sent concurrently.
		\end{itemize}
		
		\subsubsection{\textsf{Lattice of Local States}}
		
			Each process $p_i$ maintains a local view of the system, which is a subset of $M$:
		
		\begin{itemize}
			\item A process's state at any moment is a \textit{down-set} $ D \subseteq M$ , meaning if $m \in D$ and $m' \le m$, then $m \in D$.
			
			\item State transition is a \textit{join} operation in the lattice of possible message subsets. When process $p_i$ receives a message from $p_j$, it updates its state to $D_i \cup D_j$, i.e., to the \textit{supremum} of the two views. 
		\end{itemize}
		
		\subsubsection{\textsf{Vector Agreement is an Order-Preserving Map}}
		
			The output vector $\Vec{V}$ is defined as an $n$-tuple $(v_1, v_2, ...v_n)$, where:
		
		\begin{itemize}
			\item The Domain: The set of all possible input values $V = (v_1, v_2, ...v_n)$.
			
			\item The Order: Let $\mathcal{P}(V)$ be the power set of inputs, ordered by inclusion $(\subseteq)$
			
			\item Formal Goal: The algorithm aims to find a \textbf{\textit{fixed point}} in the global lattice where all non-faulty processes agree on the same subset of $\mathcal{P}(V)$.
		\end{itemize}

	\subsection{\textsf{Deterministic Consensus with Vector Agreement}}	
	
		The objectives such consensus must accomplish are:
	
		\begin{itemize}
			\item Total Order from Partial Order.
			While message delivery is only a Partial Order, the rules for updating the individual vectors $\Vec{V_i}$ are \textit{strictly monotonic}. This ensures that as messages are added to the \textit{down-set}, the vector of agreed-upon values: i) can only grow; ii) never shrink or change; and iii) \textit{eventually} reach a state where $\Vec{V}$ is identical for all processes. Objective One: transform the \textit{eventual} into \textit{guaranteed}.
		
			\item Deterministic Convergence
			In order-theoretic terms, the system is modelled as a \textit{bounded join-semilattice}. The set of \textit{initial inputs} is finite and the transition function $\Phi$ is a \textit{monotonic join-endomorphism}. Objective Two: reach a stable state, regardless of the asynchronous delays between \textit{joins} and the possible crashes.
		\end{itemize}

	\subsection{\textsf{No Theoretical Guarantee}}	
	
		A \textit{join‑endomorphic} function maps a mathematical structure – such as the lattice of possible process states – back into itself. Because the join operation is associative and commutative, the resulting \textit{supremum} is guaranteed to be identical regardless of the order in which messages are delivered. However, this guarantee holds only in the absence of unannounced process crashes.
	
		The lattice is finite, and the transition function $\Phi$ is monotonic. Therefore, when no crashes occur, the system operates with \textit{guaranteed termination}, meaning it must eventually reach a \textit{fixed point}. This fixed point corresponds to \textit{vector agreement}, where all non‑faulty processes converge on an identical view of the initial inputs, thereby ensuring safety. However, no such theoretical guarantee exists when process crashes are possible.

	\subsection{\textsf{Stable Fixed Point on a Join-Semilattice}}
	
		We define the consensus problem as the task of identifying a \textit{stable fixed point} within a \textit{complete join‑semilattice}. A consensus algorithm achieves this by ensuring monotonic convergence toward a unique supremum.
	
	\subsubsection{\textsf{System State Space as a Power Set Lattice}}
	
		Let $I = {v_1, v_2, ..., v_n}$ be the set of \textit{initial inputs}. The state of any process $p$ is represented as a vector $X_p \in \mathcal{L}$, where $\mathcal{L}$ is the lattice, defined as $\mathcal{L} = (\mathcal{P}(I) \cup {\perp})^n$, where $\mathcal{P}(I)$ is the power set of \textit{initial inputs}. The lattice $\mathcal{L}$ is partially ordered by component-wise inclusion $(\subseteq)$. For two states $X, Y \in \mathcal{L}$ , we say $X \le Y$ if every set in $X$ is a subset of the corresponding set in $Y$.
	
	\subsubsection{\textsf{Vector Agreement as Highest Fixed Point}}
	
		A consensus algorithm is defined by its transition function $\Phi: \mathcal{L} \rightarrow \mathcal{L'}$. According to the Knaster-Tarski theorem \cite{Tarski_1955}, any monotonic function on a complete lattice has a fixed point. 
		
		In an asynchronous system with possible crashes, however, processes $p_i$ and $p_j$ may reach their individual fixed points at different heights of the lattice, leading to  $\Vec{V_i} \ne \Vec{V_j}$. This divergence is avoided only when both processes reach their fixed points at the \textit{highest reachable height} of the lattice – thus ensuring that their views coincide.
	
	\subsubsection{\textsf{Possibility of Consensus}}
	
		Reaching the \textit{highest reachable height} in the lattice for all correct processes is both a necessary and sufficient condition for achieving fault‑tolerant consensus with \textit{any} agreement. Demonstrating ability to satisfy this condition with a particular consensus algorithm ensures that this algorithm can guarantee operation with both safety and liveness.
		
		The next section demonstrates, in detail, how a message‑passing consensus system can satisfy this condition using event‑based coordination alone.
		
	\section{\textsf{THE SOLUTION}}	
	
	\subsection{\textsf{Overview}}
	
		The presented here crash-tolerant consensus algorithm solves in total asynshrony the \textit{vector agreement} problem in three phases with exchange of three types of messages. 
	
	\subsubsection{\textsf{Initial Phase}}
	
		A process $p_i$ starts the algorithm by entering Initial Phase with a broadcast of an \textit{initial input} message. Each received \textit{initial input} moves $p_i$ one step up on the lattice. On knowing the \textit{initial inputs} of $(n-f)$ processes including itself, $p_i$ completes this phase and enters Proposals Phase by broadcasting an \textit{agreement proposal} message.
	
	\subsubsection{\textsf{Proposals Phase}}
	
		During the Proposals Phase, the processing of \textit{initial inputs} continues uninterrupted. When a process $p_i$ receives an initial input message that complements its most recent \textit{agreement proposal}, it moves one step upward on the lattice. When it receives an agreement proposal message that complements its own latest proposal with one or more previously unknown initial inputs, it moves one or more steps upward accordingly.
	
		Upon reaching a new lattice height, excluding any heights it skipped,  $p_i$ multicasts a new agreement proposal message. Therefore, each received agreement proposal reflects the lattice height of the process that originated it. This allows the recipient process to determine whether a majority of processes has reached a fixed point. Once $p_i$ observes $(n-f)$ identical proposals from a majority of $(n-f)$ processes, including itself, it completes this phase – knowing that the majority of processes reached a fixed point.
	
		In an asynchronous setting, detecting that a majority of processes has broadcast identical proposals becomes a potential threat to safety. The accumulation of proposals is neither commutative nor associative with respect to reaching the completion threshold. Due to asynchrony, two processes may accumulate their sets of proposals, each indicating that a majority has reached a fixed point, at different lattice heights.
	
		The lattice height at which $(n-f)$ processes broadcast the proposals accumulated by $p_i$ as evidence of reached threshold determines the content of its \textit{decision seed}. Then $p_i$ broadcasts it to complete the Proposals Phase and enter the Decision Phase.
	
	\subsubsection{\textsf{Decision Phase}}
	
		The Decision Phase enables the transition function $\Phi$ to perform the final step toward agreement. Its purpose is to prevent processes from terminating with different decision vectors. In this phase, the system identifies the highest fixed point and ensures that all correct processes adopt the view associated with that point as the agreement value.
	
		During this phase, when a process $p_i$ receives a \textit{decision seed} that complements its own broadcast decision seed with one or more previously unknown initial inputs, it moves upward on the lattice. Once $p_i$ receives $(n-f-1)$ decision seeds and has moved to the highest lattice height occupied by a known seed, the algorithm terminates.
	
		The resulting vector agreement value is the view at the \textit{highest fixed point} of the consensus-system lattice. A \textit{deterministic consensus algorithm} is therefore one that guarantees a monotonic transition of all correct processes toward this point. For any other form of agreement, processes compute the final value locally using the agreed vector as input, being confident in the global atomic consistency \cite{Herlihy_Wing_1990} of the results.

	\subsection{\textsf{Deeper in the Algorithm Phases}}

		Here we present the algorithm in a more detailed and more formalised manner. The exposition emphasises the lattice height $H_i = H(V_i) = |V_i|$ that process $p_i$ attains as it ascends through the steps of each phase, as well as the three aspects of convergence at each completion threshold:

		i) the satisfied conditions,

		ii) what these conditions indicate, and

		iii) the degree of established order they guarantee.
		
	\subsubsection{\textsf{Initial Phase}}

		Process $p_i$ starts this phase by broadcasting an \textit{initial input} message. While being in it, $p_i$ receives and processes the \textit{initial input} message of the other processes. By broadcasting its own message, it positions itself on the Step One of climbing the lattice. The steps of climbing the lattice by $p_i$ are:
		
		\begin{itemize}
			\item \textit{Step One}. Lattice height $H(V_{i, Step=1}^{Init}) = 1$, where $V_{i, Step=1}^{Init}$ is the view of $p_i$ on Step One. 
			\item \textit{Step Two}. Lattice height $H(V_{i, Step=2}^{Init}) = 2$.
			\item \textbf{. . .}
			\item \textit{Step (n-f)}. Lattice height $H(V_{i, Step=(n-f)}^{Init}) = (n-f)$.
		\end{itemize}
		
		Completion conditions:
		
		\begin{itemize}
			\item \textit{Threshold}: $p_i$ reached lattice height $(n-f)$.
			
			\item \textit{Ensured}: $p_i$ knows a subset of \textit{initial inputs} $S_i$ from $(n-f)$ processes, including its own.
		\end{itemize}
		
		Reaching the completion threshold indicates:
		
		\begin{itemize}
			\item \textit{Meaning for $p_i$}:
			\begin{itemize}
				\item "If I wait for more \textit{initial inputs} I risk \textit{infinite} waiting."
				\item "I have accumulated enough \textit{initial inputs} to propose an agreement value."
			\end{itemize}
			\item \textit{Obligation}: Process $p_i$ must complete this phase and enter the Proposals Phase by broadcasting an \textit{agreement proposal} message containing $V_{i, Step=(n-f)}^{Init}$.
		\end{itemize}
		
	\subsubsection{\textsf{Proposals Phase}}

		Process $p_i$ starts this phase by broadcasting an \textit{agreement proposal} message. An agreement proposal contains all known initial inputs, i.e., a vector of $n$ elements where $f$ elements contain a $\perp$ value.
		In this phase, $p_i$ continues receiving and processing \textit{initial input} messages, and starts processing the earlier received and newly received \textit{agreement proposal} messages. During this phase $p_i$ move up $\le f$ lattice height levels. The steps of climbing the lattice by $p_i$ are:

	\begin{itemize}
		\item \textit{Step Zero}. Lattice height $H(V_{i, Step=0}^{Prop}) = ((n-f)+0)$, where $V_{i, Step=0}^{Prop}$ is the view of $p_i$. 
		\item \textit{Step One}.  Lattice height $H(V_{i, Step=1}^{Prop}) = ((n-f)+1)$.
		\item \textit{Step Two}.  Lattice height $H(V_{i, Step=2}^{Prop}) = ((n-f)+2)$.
		\item \textbf{. . .}
		\item \textit{Step (f)}.  Lattice height $H(V_{i, Step=f}^{Prop}) = ((n-f)+f)$.
	\end{itemize}

		On every new level on the lattice, except the level(s) skipped due to a \textit{proposal} message that causes $p_i$ to jump over one or more steps, $p_i$ broadcasts a message containing its current view $V_i$. Thus $p_i$ shares with the rest of the processes every update of its view and keeps the view contained in every received \textit{proposal} message. Completion conditions:

	\begin{itemize}
		\item \textit{Threshold}: Process $p_i$ accumulated $(n-f)$ equal views from $(n-f)$ processes, containing $(n-f)$ \textit{initial inputs}, including the last broadcast view of $p_i$.
	
		\item \textit{Ensured}: Existence of a quorum set $S^{Quor}$ comprising a quorum number of $(n-f)$ processes having equal views, each containing $\ge (n-f)$ \textit{initial inputs}.
	\end{itemize}

		Reaching the completion threshold indicates:

	\begin{itemize}
		\item \textit{Meaning for $p_i$}: "I know that every process in the quorum set $S^{Quor}$ has the same view $V^{Quor}$, about a partial order of \textit{initial inputs,} and the $V^{Quor}$ view \textit{could be} the decision of the full set of correct processes $S$.
	
		\item \textit{Obligation}: Process $p_i$ must complete this phase and enter the Decision Phase by broadcasting a \textit{decision seed} message containing $V^{Quor}$.
	\end{itemize}

		The relationships between two processes $p_i$ and $p_j$, $i\ne j$, their respective vies $V_i$ and $V_j$, and the respectively contained by $V_i$ and $V_j$ sets of \textit{initial inputs} $S_i$ and $S_j$ is described as:

	\begin{itemize}
		\item $H(V_i) = H(V_j) \nrightarrow S_i = S_j$
	
		\item $S_i = S_j \rightarrow H(V_i) = H(V_j)$ 
	
		\item $p_i$ and $p_j$ may complete with $|V_i| \ne |V_j|$.
	
		\item Completion with $|V_i|=|V_j| \rightarrow S_i = S_j$
	
		\item Completion with $|V_i|<|V_j| \rightarrow S_i \subset S_j$
	
		\item Completion with $|V_j| \ge (n-f) \rightarrow |V_i|<|V_j|$
	
	\end{itemize}	

	During this phase, the consensus system faces two possible threats that are likely to jeopardise the safety of results:
	
	\begin{itemize}

	\item \textbf{Safety Threat One}: $H(V_i) > H^{Max}(V^{Quor})$. 
		\textit{Description}: 
		The majority of processes move up on the lattice until they reach a height $H^{Max}$ where their views $V^{Quor}$ are equal. If the process $p_i$ has reached a higher height on the lattice than the $H^{Max}$, this phase enforces it to ignore its "extra" data. This is not a deletion. It is a logical filter that eliminates this set of the possibilities that otherwise would have spoiled the safety of consensus.

\vspace{\baselineskip}

	\item \textbf{Safety Threat Two}:  $H(V_i^{Quor}) < H^{Max}(V^{Quor})$. 
		\textit{Description}: 
		Process $p_i$ may accumulate a quorum set $S_i^{Quor}$ at height $H(V_i^{Quor})$ that may be lower on the lattice than the height of the current views of the processes in $S_i^{Quor}$. The $V_i^{Quor}$ views could have been broadcast at any height below the $H^{Max}$.
	\end{itemize}

		The transition function $\Phi$ eliminates the Safety Thread One in the Proposals Phase of the algorithm. Yet $\Phi$ cannot eliminate the Safety Threat Two in this phase, i.e. with the exchange of \textit{agreement proposals}. Threat Two is eliminated during the Decision Phase with exchange of \textit{decision seed} messages.

	\subsubsection{\textsf{Decision Phase}}
	
		Process $p_i$ begins this phase by broadcasting a decision seed message containing the vector $V_i^{Quor}$. Once $p_i$ enters the Decision Phase, it stops processing any subsequently received \textit{initial input} or \textit{agreement proposal} messages and processes only incoming \textit{decision seed} messages.
	
		Process $p_i$ may have completed the Proposals Phase by moving a few steps down on the lattice as a consequence of how the Proposals Phase eliminates Threat One or as a consequence of the inability of the Proposals Phase to eliminate Threat Two. The purpose of the Decision Phase is to eliminate Threat Two. During this elimination, $p_i$ may move one or more steps upward on the lattice.
	
		The number of upward steps depends on the lattice height at which $p_i$ entered the Decision Phase and the height at which the majority of processes entered it. Once these adjustments converge, the Decision Phase ensures that all correct processes align on the same final view. The full diapason of steps, $p_i$ might have to climb when $H(V_i^{Quor}) < H^{Max}$ is:

\begin{itemize}
	\item \textit{Step Zero}. Lattice height $H(V_{i, Step=0}^{Seed/Term}) = ((n-f)+0)$. 
	\item \textit{Step One}.  Lattice height $H(V_{i, Step=1}^{Seed/Term}) = ((n-f)+1)$.
	\item \textit{Step Two}.  Lattice height $H(V_{i, Step=2}^{Seed/Term}) = ((n-f)+2)$.
	\item \textbf{. . .}
	\item \textit{Step (f)}.  Lattice height $H(V_{i, Step=f}^{Seed/Term}) = ((n-f)+f)$.
\end{itemize}

	Every process $p_i$ enters this phase with a broadcast of $ V_{i}^{Seed}$ containing $V_i^{Quor}$ vector. On receiving $(n-f-1)$ messages, each containing $ V^{Seed}$ equal to $ V_{i}^{Seed}$, the transition function $\Phi$ confidently uses the value of vector $ V_{i}^{Seed}$ as the agreement value for termination $ V_{i}^{Term}$. Process $p_i$ may receive one or more messages containing vector $ V_{j}^{Seed}$, where $H(V_{i}^{Seed}) < H(V_{j}^{Seed})$ making it to climb one or more steps up before it receives $(n-f-1)$ messages, each containing $ V^{Seed}$ equal to $ V_{i}^{Seed}$. Important is that when this happens, $\Phi$ confidently terminates knowing with mathematical certainty that every correct process will terminate with exactly the same $ V^{Term}$.

	Phase's completion conditions:

	\begin{itemize}
		\item \textit{Threshold}: Process $p_i$ accumulated $(n-f)$ equal \textit{decision seed} messages, including its own.
	
		\item \textit{Ensured}: $\forall p_k \in \Pi_{correct}, V_{k}^{Term} = V_{i}^{Term}$.
	\end{itemize}

	Reaching the completion threshold indicates:

	\begin{itemize}
		\item \textit{Meaning for $p_i$}: 
		\begin{itemize}
			\item "I know that the Proposals Phase ensures that if process $p_k \in \Pi_{correct}$ starts the Decision Phase with $H(V_{k}^{Seed}) < H^{Max}$, where $H^{Max}$ is the highest height on the lattice of $\ge (n-f)$ processes starting it with $V_{Majority}^{Quor}$, then $V_{k}^{Seed} \subset V_{Majority}^{Quor}$".
			\item "I know that the Decision Phase ensures that if $V_{k}^{Seed} \subset V_{Majority}^{Quor}$", then $p_k$ moves up on the lattice until $H_k = H^{Max}$ and $V_{k}^{Term} = V_{Majority}^{Quor}$."
			\item "I can complete this phase and terminate the algorithm with $V_{i}^{Term}$as agreement vector".		
		\end{itemize}
	
		\item \textit{Obligation}: $p_i$ must terminate the algorithm. 
	\end{itemize}

	\subsection{\textsf{Algorithm Rules}}
	
	\subsubsection{Order Rule}
	
		\textit{Definition}: Order Rule. The handling order of the received messages is in the order of sending and not before the phase related to them.
	
		\textit{Effect}: It neutralizes two effects of asynchrony: i) processes do not start a round of the algorithm simultaneously; and ii) messages may be received not in the order of sending.
	
		\textit{Example}: While in the Initial Phase, process $p_j$ receives an \textit{agreement proposal} message from process $p_i$ containing $(V_{i, Step=1}^{Prop}$ and later an \textit{agreement proposal} message of process $p_i$ containing $(V_{i, Step=0}^{Prop}$. Process $p_j$ records both messages in the memory and handles them after entering the Proposals Phase. $p_j$ processes the message containing $V_{i, Step=0}^{Prop}$ first and the message containing $V_{i, Step=1}^{Prop}$ second.
	
		\textit{Requirement}: If process $p_j$ receives an \textit{agreement proposal} message from process $p_i$ containing $(V_{i, Step=2}^{Prop}$ and later an \textit{agreement proposal} message of process $p_i$ containing $(V_{i, Step=0}^{Prop}$. Process $p_j$ does not know whether process $p_i$ has broadcast an agreement proposal message on Step One before broadcasting one on Step Two or $p_i$ skipped over Step One and has broadcast an agreement proposal message only on Step Two. To avoid the possible ambiguity, every broadcast \textit{agreement proposal} message must contain the sequentially ordered history of all broadcasts so far by the same process \textit{agreement proposal} messages.
		
	\subsubsection{Blend Rules}
	
		\textit{Effect}: Enforce indirect distribution of one or more \textit{initial values}.
	
		\textit{Note}: In contrast, reliable broadcast sends messages indirectly pursuing different objectives.
		
	\vspace{\baselineskip}
	
		\textit{Definition}: Blend Rule 1. If process $p_j$ receives an \textit{agreement proposal} message from process $p_i$ containing $V_{i}^{Prop}$ where $H(V_{i}^{Prop}) \le H(V_{j, Last}^{Prop})$, where $V_{j, Last}^{Prop}$ is the content of the last broadcast by $p_j$ \textit{agreement proposal}, and $V_{i}^{Prop}$ complements it with an unknown \textit{initial input}, process $p_j$ must update its $V_{j}^{Prop}$ and broadcast it with an \textit{agreement proposal} message.
	
	\vspace{\baselineskip}
	
		\textit{Definition}: Blend Rule 2. If process $p_j$ receives an \textit{agreement proposal} message from process $p_i$ containing $V_{i}^{Prop}$ where $H(V_{i}^{Prop}) > H(V_{j, Last}^{Prop})$, process $p_j$ unconditionally updates its $V_{j}^{Prop}$ by including the unknown initial input(s) and broadcast it with an \textit{agreement proposal} message.
		
	\subsubsection{Completion Rule}
	
		\textit{Definition}: Completion Rule 1. Process $p_i$ completes each phase of the algorithm on reaching the phase's completion threshold.
	
		\textit{Note}: Completion of the Decision Phase terminates the algorithm.

	\subsection{\textsf{Safety Threats Exemplified}}
	
		Consider a system of 3 processes, which tolerates one crash-fail. Every process starts the Proposals Phase by broadcasting \textit{agreement proposal} vector containing two \textit{initial inputs} and one empty vector element, denoted as $\varnothing$.
	
		Let the configuration of the \textit{agreement proposal} vectors broadcast by processes $p_1$, $p_2$, and $p_3$ is as follows:
	
	\vspace{\baselineskip}
	
		Vector broadcast by $p_1$ : $V_1^{Frst}$ = $( v_1, v_2, \varnothing )$.
	
		Vector broadcast by $p_2$ : $V_2^{Frst}$ = $( v_1, v_2, \varnothing )$.
	
		Vector broadcast by $p_3$ : $V_3^{Frst}$ = $( \varnothing, v_2, v_3 )$.
	
	\subsubsection{\textsf{Proposals Phase Illustrated}}
	
		On receiving one \textit{agreement proposal}, the receiver process $p_i$ will either broadcast a second \textit{agreement proposal} vector $V_i^{Scnd}$ and climb one step up on the lattice, or complete this phase by broadcasting a \textit{decision seed} vector, depending on which \textit{agreement proposal} is received first. For example, if $p_1$ receives $V_3^{Frst}$ before $V_2^{Frst}$, it will broadcast $V_1^{Scnd}$ = $( v_1, v_2, v_3 )$ and climb one step up. On receiving $V_2^{Frst}$ before $V_3^{Frst}$, $p_1$ will complete this phase by broadcasting vector $( v_1, v_2, \varnothing )$ as decision seed.
	
		On receiving one more \textit{agreement proposal}, the receiver process $p_i$ will either do nothing or will complete this phase. In the example with $p_1$, if it receives $V_3^{Frst}$ before $V_2^{Frst}$, it will broadcast $V_1^{Scnd}$ = $( v_1, v_2, v_3 )$, so on receiving of $V_2^{Frst}$ it will do nothing. 
	
		With process $p_3$ the sequence of receiving obviously does not matter: in both cases it will broadcast a second \textit{agreement proposal} vector $V_3^{Scnd}$ and climb one step up on the lattice. Let $p_3$ has broadcast $V_3^{Scnd}$ = $( v_1, v_2, v_3 )$ as response to received $V_1^{Frst}$. In this case if $p_3$ receives $V_2^{Frst}$ before $V_1^{Scnd}$, it will complete this phase by broadcasting $( v_1, v_2, \varnothing )$ as decision seed. On receiving $V_1^{Scnd}$ before $V_2^{Frst}$, $p_3$ will complete this phase by broadcasting $( v_1, v_2, v_3 )$ as decision seed. 
	
		We have just illustrated why for the algorithms that operate with events-based synchronisation the time does not matted. What matters is only the receiving order.	
		
	\subsubsection{\textsf{Safety Threat One}}	
		
		Consider a scenario: i) $p_1$ receives $V_2^{Frst}$ before $V_3^{Frst}$ and completes with $( v_1, v_2, \varnothing )$; ii) $p_2$ receives $V_1^{Frst}$ before $V_3^{Frst}$ and completes with $( v_1, v_2, \varnothing )$; iii) on receiving any $V_i^{Frst}$, $p_3$ broadcasts $V_3^{Scnd}$ = $( v_1, v_2, v_3 )$ and climbs one step up on the lattice, but on receiving also a $V_j^{Frst}, j \ne i$, i.e., on becoming aware that the majority completes with $( v_1, v_2, \varnothing )$, $p_3$ does the same. It shows that the Proposals Phase handles this threat.
		
	\subsubsection{\textsf{Safety Threat Two}}	

		Consider a scenario: i) $p_1$ receives $V_3^{Frst}$ before $V_2^{Frst}$, climbs one step up on the lattice, and broadcast $V_1^{Scnd}$ = $( v_1, v_2, v_3 )$; ii) $p_2$ receives $V_3^{Frst}$ before $V_1^{Frst}$, climbs one step up on the lattice, and broadcast $V_2^{Scnd}$ = $( v_1, v_2, v_3 )$; iii) $p_3$ receives one $V_i^{Frst}$ = $( v_1, v_2, \varnothing )$, which triggers a broadcast of $V_3^{Scnd}$ = $( v_1, v_2, v_3 )$ and another $V_j^{Frst}$ = $( v_1, v_2, \varnothing )$, which together obligate $p_3$ to complete this phase by broadcasting $( v_1, v_2, \varnothing )$ as decision seed; iv) $p_1$ receives $V_2^{Scnd}$ and completse with a $( v_1, v_2, v_3 )$ as decision seed; v) $p_2$ receives $V_1^{Scnd}$ and completes with a $( v_1, v_2, v_3 )$ as decision seed. This scenario shows that the Proposals Phase cannot handle this threat. Yet the outcomes of this phase guarantee that the Decision Phase does handle it.
		
	\section{PROOFS}	
		
	\subsection{\textsf{Termination with Valid Vector Agreement}}
		
		\textbf{Theorem 1}: \textit{In a \textbf{fully asynchronous} message‑passing system with \textbf{crash‑prone} processes, a \textbf{deterministic} algorithm can achieve \textbf{total‑order broadcast} while preserving both \textbf{safety} and \textbf{liveness}, assuming that no more than a minority of processes fail.}
		
	\begin{proof}
			
		Sufficient is to demonstrate the ability to terminate reaching a decision configuration with only one decision state. The proof follows from these 4 lemmas:
			
	\end{proof}
		
	\subsection{\textsf{Vectors with Equal Cardinality Contain the Same Set of Initial Inputs}}	
		
		\textbf{Lemma 1}: \textit{Vectors of processes completing Proposals Phase with equal number of initial inputs contain the same set of initial inputs, i.e., $|V_i|=|V_j| \rightarrow S_i = S_j$.}
	
	\begin{proof}
		Completion of Proposals Phase by process $p_i$ is triggered by reaching a threshold of $(n-f)$ accumulated equal views from $(n-f)$ processes, containing $\ge (n-f)$ \textit{initial inputs}. 
		
		Let us assume that $|V_i|=|V_j|$ and $|S_i| \ne  |S_j|$. If $p_i$ has accumulated  $(n-f)$ equal views at lattice height $H(V_i)$ and $H(V_j) = H(V_i)$, then $S_j$ cannot accumulate $(n-f)$ equal views based on the following reason. No process broadcasts two different views at the same height of the lattice. Hence, this assumption is incorrect.
		
	\end{proof}
		
	\subsection{\textsf{No Process Completes Above the Majority's Highest Reachable Height}}
		
		\textbf{Lemma 2}: \textit{Majority of processes complete Proposals Phase at the highest reachable lattice height.}
	
	\begin{proof}
		After the majority of processes reach the highest reachable height and complete, no other majority can be formed at higher height.
		
	\end{proof}
	
	\subsection{\textsf{Vector with Smaller Cardinality Contains a Subset of Initial Inputs}}
		
		\textbf{Lemma 3}: \textit{Completion of Proposals Phase with $|V_i|<|V_j| \rightarrow S_i \subset S_j$}.
	
	\begin{proof}
		
		Process $p_k$ moves up 1 step higher on the lattice by adding 1 unknown \textit{initial input} $I_m$ to its view $V_k$, i.e., $V_k^{StepX+1} = V_k^{StepX} \cup I_m$. Thus for instance, process $p_i$ and $p_j$ may complete Proposals Phase with $S_i \ne S_j$ under the following sequence of events: 
		
		
		i) $p_i$ and $p_j$ have accumulated $(n-f-1)$ equal views and $V_i^{StepX}$ = $V_j^{StepX}$;  
		
		ii) $p_j$ adds 1 unknown \textit{initial input} $I_m$ to its view $V_j^{StepX}$ and moves up 1 step higher on the lattice to its new view $V_j^{StepX+1}$; 
		
		iii) $p_j$ accumulates $(n-f)$ views equal to $V_j^{StepX+1}$ and completes Proposals Phase; 
		
		iv) $p_i$ receives a message with a view equal to $V_i^{StepX}$ and accumulates $(n-f)$ views equal to $V_i^{StepX}$, which requires $p_i$ to complete with $V_i^{StepX}$, hence $S_i \subset S_j$. 
		
		\vspace{\baselineskip}
		
		To complete with $S_i \not\subset S_j$, while at lattice height X $p_i$ must have received $(n-f)$ equal views $V^{Quor} \ne V_i^{StepX}$ from processes at height X. 
		
		In this case $p_j$ would not have completed with $V_j^{StepX+1}$. Process $p_j$ could not have accumulated $(n-f)$ views equal to $V_j^{StepX+1}$ if $(n-f)$ processes have equal $V^{Quor} \ne V_j^{StepX}$. 
		
		Once a majority of $(n-f)$ processes with equal $V_q$ views is established, all processes on the same or higher height complete with $V_q$ [Lemma 2]. Hence, $S_i \not\subset S_j \rightarrow S_i = S_j$.
		
	\end{proof}

	\subsection{\textsf{The Agreed Vector Value}}
		
		\textbf{Lemma 4}: The view $V^{Quor}$ at highest reachable height $H^{Max}$ by majority of processes is the agreement value.
	
	\begin{proof}
		Decision seeds from the lower levels of the lattice are subsets of the highest [Lemma 3]. During the Decision Phase every correct process at a lower level will receive $\ge 1$ message containing the decision seed of the majority even when all of the following conditions are simultaneously presented:
		\begin{itemize}
			\item $n=2f+1$; 
			\item only $(n-f)$ processes completed Proposals Phase at the highest height; 
			\item all $f$ crashes happened immediately before broadcasting a \textit{decision seed};  
			\item all $f$ crashed processes are from the majority.
		\end{itemize}
	\end{proof}

	\section{DISCUSSION}
	
		We have demonstrated the possibility of crash‑tolerant consensus in an asynchronous, wait‑free, read/write shared‑memory model. The result is established in a strictly formal manner using the structure of a \textit{monotonic join‑semilattice}.
		Formalisms of this kind typically abstract away failures and focus on the evolution of the global state. Nevertheless, our work shows how failures can be incorporated by viewing them through the lens of \textit{state convergence} of the evolving global structure of the lattice.

	\subsection{\textsf{Reconciling with the Reconfirmation of Impossibility}}

		The proof of possibility we presented is global in the sense that it reasons about the system’s global state, in contrast to local proofs that reason from the perspective of individual process states. In this subsection, we compare and contrast our global proof of possibility with the reconfirmation of impossibility obtained through global‑state argumentation by Attiya et al.
		In both our work and that of Attiya et al., the first phase consists of processes broadcasting their \textit{initial input} values, and the second phase consists of broadcasting their \textit{views of the global state}. In both settings, exchanging views of the global state amounts to exchanging \textit{vectors} of known initial inputs, which may later be used to compute a binary agreement value.
	
	\subsubsection{\textsf{First Difference: The Negligible One}}
	
		Attiya et al. view consensus on a binary value as the most basic form of agreement. Yet, instead of relying on the valency argument, their impossibility result is established through a different mechanism: they show that termination with different binary results arises as a consequence of computing from different sets of initial input values.

		We view consensus on a vector of \textit{initial inputs} as the most fundamental one. This distinction is negligible as in both works, implicitly in theirs and explicitly in ours, the core focus is on agreement about the set of initial input values from which the final decision is computed. Whether the agreement is data‑dependent or data‑independent becomes irrelevant once the valency argument is no longer relied on.
		
	\subsubsection{\textsf{Second Difference: Further Exploring of Phase Two}}
		
		Attiya et al. observed that, due to asynchrony and the possibility of faults, processes may terminate a two‑phase protocol with binary values computed from different sets of initial inputs. This divergence can lead to termination with different binary outcomes.
		
		Our work goes further. We pursue the deeper question of why Phase Two, the Proposals Phase of our algorithm, can be completed with different sets of initial inputs, and whether this divergence can be overcome. Using the structure of a monotonic join‑semilattice as an algebraic abstraction of a lock‑free shared‑memory model, our analysis reveals the underlying cause.
		
		A direct consequence of asynchrony is the unpredictable, effectively random order in which consensus protocol messages are received. In a system of $n$ processes that tolerated $f$ crashes, no process can afford to wait for a particular type of message from more than $(n-f-1)$ other processes without risking loss of liveness through unbounded waiting. Therefore, no process waits beyond this threshold.

		All processes begin Phase Two, i.e., the exchange of individual views of the system’s global state, at the same lattice height of $(n-f)$. Consequently, the first broadcast of global state views always contains the initial inputs of $(n-f)$ processes, though these sets may not be identical. During this phase, processes continue their monotonic ascent on the lattice. Yet, because of asynchrony, each process ascends at different times and in different orders, leading to divergent views of the global state.
		
		To avoid entering a loop of infinite waiting, every process completes this phase under a strict rule: on accumulation of $(n-f)$ equal views, \textbf{including} its last broadcast view if \textbf{it is} the same, or \textbf{excluding} its last broadcast view if \textbf{it is not} the same. As shown earlier, enforcing this rule can lead to undesirable consequences: the specific threats to consensus safety.
		
		The \textit{Safety Threat One} scenario occurs when a process climbs the lattice to a height higher than the height at which the majority of processes complete Phase Two. The cause explanation is straightforward. A slow process $p_s$ may receive all views broadcast by the rest of the system, yet none of the majority processes receive the initial input or the views broadcast by $p_s$. This threat is successfully mitigated within Phase Two.
		
		The \textit{Safety Threat Two} causes a process to complete Phase Two at a lattice height below the height where the majority of processes completes, in a more subtle scenario. While processes ascend the lattice, a process $p_i$ may accumulate $(n-f)$ identical views excluding its own. From its perspective, all indicators suggest that it must handle \textit{Safety Threat One}, and it completes Phase Two accordingly. However, this leads $p_i$ to completion of Phase Two with a set of initial inputs that differs from the set used by the majority of processes. This divergence explains the impossibility of achieving crash‑tolerant consensus in an asynchronous system using a \textit{two‑phase} protocol.	
	
	\subsubsection{\textsf{Third Difference: The Three-Phase Protocol}}
		
		As a consequence of Safety Threat Two, no two‑phase protocol can solve crash‑tolerant consensus in an asynchronous setting. However, when Phase Two implements the transition function $\Phi$ in the same manner as our Proposals Phase, it establishes exactly the structural conditions required for a third phase to complete the task successfully. In particular, the Proposals Phase guarantees that, upon its completion:
			
	\begin{itemize}	
		\item Completion vectors of equal cardinality contain the same set of initial inputs [Lemma 1].
		
		\item No completion vector has cardinality greater than that of the completion vector held by a majority of the system’s processes [Lemma 2].
	
		\item A completion vector of smaller cardinality contains a subset of the initial inputs contained in any completion vector of larger cardinality [Lemma 3].
	\end{itemize}		
	
		The Third Phase, when instantiated as our Decision Phase, leverages this set of guaranteed prerequisites to reach a decision in a single messaging round [Lemma 4]. That single round is precisely the boundary between possibility and impossibility.
				
	\subsubsection{\textsf{Reconciliation of Outcomes}}
	
		The possibility of crash‑tolerant consensus in total asynchrony and its impossibility under the same model are mutually exclusive claims. Yet there is no contradiction in the intermediate reasoning steps that lead to these opposing conclusions.
		
		The impossibility to guarantee safety solely by accumulating information about the system’s global state is inherent to asynchrony when combined with the possibility of crashes. As shown earlier, \textit{Safety Threat Two} is unavoidable in Phase Two.
			
		However, a closer analysis of Phase Two reveals that its correct implementation satisfies two essential properties: \textit{subset monotonicity} and \textit{majority alignment}. These properties create the structural conditions that make safe termination achievable. Only a single messaging round is required to neutralize the consequences of this safety threat.

	\subsection{\textsf{Global Proof of Possibility}}	

		During the Proposals Phase, processes accumulate and interpret data about global state in a way that enables safe termination after only one additional messaging round. This opportunity exists because, although different processes may complete the phase with different completion vectors, the structure of these vectors is far from arbitrary. Their ordering satisfies two critical properties:

	\begin{itemize}	
		\item \textbf{Subset monotonicity}. The set of initial input values in any completion vector with fewer non‑empty entries is a subset of the set of initial input values contained in every completion vector with more non‑empty entries [Lemma 3]. 
		
		\item \textbf{Majority alignment}. A majority of processes complete the Proposals Phase with the vector containing the largest set of initial input values [Lemma 2]..
	\end{itemize}	

		Together, these properties ensure that termination and safety can be achieved with just one additional phase consisting of a single messaging round.

		\textbf{Key insight}: The transition function is aware that at least one message carrying the largest set of initial inputs will reach every correct process [Lemma 4]. With this knowledge, each process confidently terminates using the completion vector containing the largest set of initial values that it receives during the Decision Phase.

	\subsection{\textsf{Conclusion}}
	
		Deterministic, crash‑tolerant consensus in a fully asynchronous setting was regarded as impossible for more than forty years, largely for two reasons:
		
	\begin{itemize}	
		 \item \textbf{First}, the inability of any \textit{one‑phase} consensus protocol to terminate with safety was interpreted and presented as absolute and unconditional impossibility with the support of the classic \textit{valency argument}, and accepted as such within the community of researchers in distributed computing. 
		 \item \textbf{Second}, the inability of any \textit{two‑phase} protocol  to terminate with safety was interpreted and presented as reconfirmation of the absolute and unconditional impossibility with the support of \textit{global state arguments}.
	\end{itemize}	
	
		Our earlier publication presented a \textit{reduction‑ad‑absurdum} example illustrating the possibility for misuse of the \textit{valency argument} and demonstrated that termination in full asynchrony can be \textit{decoupled} from the act of making a data‑dependent decision. Any data‑dependent consensus is a \textit{design choice}: it is an artificially over‑constrained problem layered on top of vector consensus with agreement about the initial inputs. Its constraints \textit{are not} nature-established physical, logical, or topological limits.
	
		The presented here work identified and described the \textit{safety threat} that causes the inability of any \textit{two‑phase} protocol to terminate with safety that was accepted by the community of researchers in distributed computing as further confirmation of absolute and unconditional impossibility with global state arguments. We demonstrated why no \textit{two‑phase} protocol can neutralise this threat and how to eliminate its consequences with a \textit{three-phase} protocol, thereby achieving termination with unconditional safety.
	
		With this work we resolved the apparent contradiction between the possibility of consensus and the reconfirmation of its impossibility. It reconciled these exact opposite conclusions by demonstrating that a single protocol phase, with only one messaging round, separates possibility and impossibility results. Our conclusions are supported and verified with a strictly formal proof, which establishes the possibility of consensus in a totally asynchronous system capable of tolerating multiple process crashes.

		The proven by other authors inability to resolve crash-tolerant consensus with one‑phase and two‑phase protocols represent essential milestones in the theoretical development of consensus in asynchrony. The presented here algorithm, together with its proof, constitutes the next milestone in this progression. As demonstrated in a strictly formal manner, the first two phases establish the structural groundwork that makes safe termination achievable, and the third phase completes the objective.

	\section*{Funding}
	
		This work is partly sponsored by the Australian Federal Government through the Research and Development Tax Incentive Scheme. 
		
	\newpage
	
	\bibliographystyle{tfnlm}
	\bibliography{CiASF}

	\end{document}